\documentclass[aps,superscriptaddress, nofootinbib, twocolumn]{revtex4}
\usepackage{amsfonts,amssymb,amsmath}            
\usepackage[]{graphics,graphicx,epsfig}            
\def\identity{\leavevmode\hbox{\small1\kern-3.8pt\normalsize1}}

\newcommand{\ket}[1]{\left | #1 \right\rangle}
\newcommand{\bra}[1]{\left \langle #1 \right |}
\newcommand{\half}{\mbox{$\textstyle \frac{1}{2}$}}
\newcommand{\smallfrac}[2][1]{\mbox{$\textstyle \frac{#1}{#2}$}}
\newcommand{\Tr}{\text{Tr}}
\newcommand{\braket}[2]{\left\langle #1|#2\right\rangle}
\newcommand{\proj}[1]{\ket{#1}\bra{#1}}
\renewcommand{\epsilon}{\varepsilon}
\begin{document}

\title{Optimal Asymmetric Quantum Cloning}
\date{\today}

\author{Alastair \surname{Kay}}
\affiliation{Centre for Quantum Technologies, National University of Singapore, 
			3 Science Drive 2, Singapore 117543}
\affiliation{Keble College, Parks Road, University of Oxford, Oxford, OX1 3PG, UK}
\author{Ravishankar \surname{Ramanathan}}
\affiliation{Centre for Quantum Technologies, National University of Singapore, 3 Science Drive 2, Singapore 117543}
\author{Dagomir \surname{Kaszlikowski}}
\affiliation{Centre for Quantum Technologies, National University of Singapore, 3 Science Drive 2, Singapore 117543}
\affiliation{Department of Physics, National University of
 Singapore, 2 Science Drive 3, Singapore 117542}

\begin{abstract}
While the no-cloning theorem, which forbids the perfect copying of quantum states, is well-known as one of the defining features of quantum mechanics, the question of how well the theory allows a state to be cloned is yet to be completely solved. In this paper, rigorous solutions to the problem of $M\rightarrow N$ asymmetric cloning of qudits are obtained in a number of interesting cases. The central result is the solution to the $1 \rightarrow N$ universal asymmetric qudit cloning problem for which the exact trade-off in the fidelities of the clones for every $N$ and $d$ is derived. Analogous results are proven for qubits when $M=N-1$. We also consider state-dependent $1 \rightarrow N$ qubit cloning, providing a general parametrization in terms of a Heisenberg star Hamiltonian. In all instances, we determine the feasibility of implementing the cloning economically, i.e., without an ancilla, and determine the dimension of the ancilla when an economic implementation is not possible.
\end{abstract}

\maketitle

\section{Introduction}

Quantum information constitutes a feasibility study of protocols subject to the physical laws of quantum mechanics. When they cannot be realized, it quantifies and determines the best possible result. Often these considerations are made under restrictions such as those of local operations and classical communication. However, even in their absence, there is the fundamental restriction that any protocol must be implemented by a completely positive (CP) map. The quintessential instance of this problem is quantum cloning; it is impossible to take one copy of an unknown quantum state and produce a second identical copy, unless that unknown state is known to be drawn from a set of orthogonal states \cite{WZ82a}.

The no-cloning theorem, as a cornerstone of quantum theory, is related to many fundamental ideas such as the principle of no-signaling \cite{Gisin98}, uncertainty relations and state estimation \cite{Gisin97}. Apart from the intrinsic theoretical interest, the no-cloning theorem has also found application in quantum cryptography \cite{BB84} where it enables detection of attempts by an adversary to copy the information on a communication channel. While the no-cloning theorem is now well established, the question of the extent to which an unknown quantum state can be copied has been the subject of intensive research, excellent reviews of which can be found in \cite{rmp, review}.

The optimal cloning of discrete quantum states began with the idea of the Buzek-Hillery quantum cloning machine which obtains two identical copies of a given unknown qubit state \cite{BH96}. This has been extended to $M \rightarrow N$ cloning \cite{MNUQCM} where, starting from $M$ copies of the same unknown quantum state, the task is to produce $N$ output copies of as high a quality as possible. While the original cloning machines were symmetric, in the sense that all output copies had the same fidelity, this has also been extended to asymmetric cloning \cite{asymmetric}. In this latter task, not all copies need to have the same quality; some output clones can be made to have higher fidelities at the expense of others. This is an important feature since it is precisely these trade-offs which quantify a potential level of eavesdropping, or the relationship between different correlations in a given quantum state.

The original cloning machines were designed to clone all input pure states of a given dimension $d$ (referred to as ``universal cloning"). With prior knowledge of the distribution of the states to be cloned, better strategies can be devised for this, state-dependent, cloning. There are several well-known instances, the most common of which we refer to as ``equatorial cloning" \cite{dariano-2003}: for $d=2$, the state of the qubit is known to be drawn from the set of states on the equator of the Bloch sphere, $(\ket{0}+e^{i\phi}\ket{1})/\sqrt{2}$ and any angle $\phi\in(0,2\pi)$ is equally likely. This generalizes to ``phase-covariant" cloning: for $d=2$, the state of the qubit is drawn from the set $\cos\theta\ket{0}+\sin\theta e^{i\phi}\ket{1}$ with a probability distribution that is independent of the parameter $\phi$.

Another important question is the economy of implementation of a cloner; if the optimal cloning machine can be implemented unitarily without any ancillary systems, the cloner is said to be economical, otherwise it is not \cite{economy, economic}. The presence or absence of any ancilla systems alters the implementation of the cloner experimentally, the economic cloner being simpler to control and less sensitive to decoherence effects. 

In this paper, we apply the powerful formalism of the Choi-Jamio\l kowski isomorphism to solve several interesting instances of the cloning problem. The central result constitutes a completion of the study started in \cite{our_paper}; a determination of the optimal cloning fidelities for $1\rightarrow N$ universal asymmetric cloning of qudits (arbitrary $N$ and $d$). While special case results are known for $N$ up to 4 \cite{asymmetric, subsume}, we proposed in \cite{our_paper} a general solution, but some caveats remained unresolved. We simplify the proof of \cite{our_paper}, and extend it to close the potential loopholes in the argument, making use of the Lieb-Mattis theorem \cite{liebmattis} to prove that the maximum eigenvalue of the whole system is found in a particular subspace.

 The paper is organized as follows. In Section \ref{sec:isomorphism}, we explain the formalism used to obtain bounds on the achievable fidelities in state transformation tasks and state the necessary and sufficient condition for the maximum fidelity to be realized by a completely positive map. We also explain the applicability of the formalism to the problem of cloning quantum states, stating the figure of merit used to define the success of the operation. In section \ref{sec:qubits}, we study the $1 \rightarrow N$ qubit cloning problem, providing a parametrization for a class of state-dependent cloners, deriving solutions in the symmetric instance and studying the economy of its implementation. In the asymmetric case, it is proven that if $N$ is odd,the optimal cloner is economical. For even $N$, no more than one ancilla qubit is required. For equatorial qubit cloning, we solve the asymmetric case for $N=2$, enabling a simple derivation of the monogamy of the CHSH Bell inequality.

Section \ref{sec:main} constitutes the main result of the paper, where the $1 \rightarrow N$ asymmetric universal qudit cloning problem is solved and its economic realization is studied. Many of the technicalities of the proof, which elevate it over the conjectures in \cite{our_paper}, are expounded in the Appendix. For $N>2$, we prove that the optimal machine is economical. For $N=2$, the optimal machine is not economical, requiring one additional qudit\footnote{other than extreme cases which reduce to $N=1$}. Subsequently, in Sec.\ \ref{sec:many}, the applicability of the formalism to the $M \rightarrow N$ cloning is explained. The penultimate section, Sec.\ \ref{sec:trades}, contains the trade-off relations in the optimal fidelities in the two instances where such a relation can be derived, namely $1 \rightarrow N$ qudit cloning and $(N-1) \rightarrow N$ qubit cloning. The paper concludes with a summary of the work and possible future extensions.

\section{The Choi-Jamio\l kowski Isomorphism} \label{sec:isomorphism}
We begin with an explanation of the main tool used in the solution of optimal cloning tasks, the well-known Choi-Jamio\l kowski isomorphism. This formalism is used in general to find how well a particular state transformation task can be achieved by a quantum process, i.e., a completely positive map \cite{iso}. 

The scenario is that, given one of a set of $L$ states $\ket{\psi_i}$ ($i = 1, \dots, L$), we are required to perform a particular state transformation on it, without being told which of the states we have been given. The required transformation may not be achievable exactly within the quantum formalism, but is best approximated within the theory by a completely positive, trace preserving map $\mathcal{E}$ that transforms input state $\ket{\psi_i}$ into $\mathcal{E}(\ket{\psi_i})$. The success of the state transformation task is then measured by a fidelity given by 
\begin{equation}
F=\frac{1}{L}\sum_i\Tr\left(\mathcal{M}_i\mathcal{E}(\ket{\psi_i})\right). \nonumber 
\end{equation}
Here $\mathcal{M}_i$ are positive operators ($\mathcal{M}_i \geq 0$) satisfying $\|\mathcal{M}_i\|\leq 1$ so that $F$ is indeed a fidelity taking values between $0$ and $1$. If the fidelity takes value $1$, we infer that the map has perfectly implemented the required state transformation for all the specified input states. As a simple example, consider being required to transform the states $\ket{\psi_i}$ into states $\ket{\phi_i}$, in which case we simply define $\mathcal{M}_i=\proj{\phi_i}$. 

The fidelity can now be rewritten in a manner that yields definite upper bounds. This is accomplished by the isomorphism as follows. Since $\mathcal{E}$ is a completely positive map, its operation on a subsystem $O$ of a bipartite entangled state (entangled between input system $I$ and output $O$) is well defined. Let us denote the maximally entangled state of interest as
$$
\ket{B}=\frac{1}{\sqrt{d}}\sum_{n=0}^{d-1}\ket{\phi_i}_I\ket{\phi_i}_O,
$$
where the basis states $\ket{\phi_i}$ ($i = 0, \ldots, d-1$) span the subspace occupied by the set of input states $\ket{\psi_i}$ (which have dimension $d$). The action of the composite map, composed of the identity on the input space and the desired map $\mathcal{E}$ on the output space of the maximally entangled state gives us the output $\chi_{IO}$,
$$
\identity_I\otimes \mathcal{E}(\ket{B})_O=\chi_{IO}.
$$
The condition that $\mathcal{E}$ be trace preserving then implies that $\Tr(\chi_{IO})=\braket{B}{B}=1$. If the map is not trace preserving, this would instead yield $\Tr(\chi_{IO})\leq 1$, which will not alter our conclusions since we are interested in finding an upper bound to the fidelity of state transformation. The fidelity can now be evaluated as 
$$
F=\frac{d}{L}\sum_i\Tr\left(\chi_{IO}(\proj{\psi_i}^T_I\otimes \mathcal{M}_i)\right),
$$
where the superscript $T$ denotes transposition. 
If we define a matrix $R$ as 
\begin{equation}
R=\frac{d}{L}\sum_i\proj{\psi_i}^T_I\otimes \mathcal{M}_i, \label{eqn:def_R}
\end{equation}
then we obtain 
\begin{equation}
F=\Tr(R\chi_{IO}) 
\leq  \Tr(\chi_{IO})r_{\max} 
\leq  r_{\max}
\end{equation}
where $r_{\max}$ is the maximum eigenvalue of matrix $R$. We thus arrive at the important conclusion that the maximum fidelity achievable within quantum theory for a given state transformation task is bounded by the maximum eigenvalue of a suitable matrix $R$.

A few points are worthy of emphasis. The dimension $d$ appearing in the equations above is the dimension of the subspace of Hilbert space spanned by the set of states $\ket{\psi_i}$. Even if the states $\ket{\psi_i}$ are defined on $M$ qubits, it is not necessarily the case that $d = 2^M$, since the states might exist on a smaller subspace. Such is indeed the case when we are performing the $M \rightarrow N$ cloning task as we shall see in a section \ref{sec:many}. For a continuous set of states, the sum appearing in the definition of $R$ (Eq.\ (\ref{eqn:def_R})) transforms to an integral. The factor of $1/L$ appearing in Eq.\ (\ref{eqn:def_R}) stems from the assumption that each of the input states $\ket{\psi_i}$ is equally likely. If this is not the case, these parameters can be adjusted based on a given probability distribution of the input states. 

The fact that the state transformation problem has been transformed into the problem of finding the maximum eigenvalue of a matrix is interesting and highly useful. Even if the matrix $R$ proves to be difficult to diagonalize exactly, one can use many techniques (borrowed for instance from condensed matter physics) to bound the achievable fidelities. One example of such a technique that has been used in finding bounds on ground state energies of condensed matter systems is to upper bound the norm of the matrix by the sum of the norms of the constituent terms \cite{anderson}. 

\subsection{Achieving the Maximum Fidelity}

From the equations above, we see that the maximum eigenvector, $\ket{\Psi}$, of the matrix $R$ defines the optimal strategy if it can be realized. If $\ket{\Psi}$ is unique, it can be expressed as a pure bipartite state between the subsystems $I$ and $O$ with a Schmidt decomposition
$$
\ket{\Psi}=\sum_{n=0}^{d-1}\beta_n\ket{\phi_n}\ket{\lambda_n}.
$$
When the Schmidt coefficients are given by $\beta_n^2=\smallfrac{d}$, the state is maximally entangled across the input/output partition and can be implemented by a unitary $U$, defined as
$$
U\ket{\phi_n}=\ket{\lambda_n} \qquad\forall n.
$$
Here the relevant Hilbert spaces are extended as necessary so that they have the same size. In this instance, the optimal strategy is called economical, meaning that one does not require an ancilla for the operation to be implemented. In fact, even if the maximum eigenvector is not unique, provided there exists a superposition of the maximum eigenvectors that is maximally entangled, the optimal map can be implemented as a unitary and is therefore economical. 

More generally, if there exists a mixture of the maximum eigenvectors of $R$, $\rho_R$, such that $\Tr_O\rho_R$ is maximally mixed, then this can be implemented as a CP map or, equivalently, a unitary operator over a larger Hilbert space, in which case the operation is no longer economical. The condition for implementation of the state transformation task optimally by a CP map is therefore 
$$
\Tr_O\rho_R = \smallfrac{d} \identity_{I}.
$$
That this condition is sufficient is seen easily by recognizing that one can add an auxiliary Hilbert space to purify $\rho_R$. The overall pure state then defines a unitary as in the previous case although this operation is not economical. That this condition is also necessary is seen by writing the optimal CP map using the Kraus decomposition as 
$$
\mathcal{E}(\rho)=\sum_i A_i\rho A_i^\dagger
$$
where $\sum_i A_i^\dagger A_i=\identity$. We then find that 
$$
\rho_R=\identity_I\otimes\mathcal{E}_O(\proj{B})=\sum_iA_i^O\proj{B}_{IO}{A_i^O}^\dagger.
$$
And upon performing the partial trace over $O$, we obtain 
$$
\Tr_O\rho_R=\Tr_O\left(\sum_iA_i^\dagger A_i \proj{B}_{IO}\right),
$$
which can be reduced using the completeness relation for the Kraus operators $\sum_iA_i^\dagger A_i=\identity$ to $
\Tr_O\rho_R=\smallfrac{d} \identity_{I}$.

As a final remark we mention that the optimal strategy can also be implemented by teleporting the input state onto spin $I$ of a resource state which could be either $\ket{\Psi}$ or the purification of $\rho_R$. The different measurement results of teleportation can be corrected for by action on the output space if the state satisfies the required trace condition. In the case of cloning, this gives rise to telecloning protocols.

\subsection{Application to cloning quantum states}

The potentially powerful formalism described above is now used in the problem of the optimal cloning of quantum states. In the cloning process, we start with an unknown quantum state $\ket{\psi}$ of Hilbert space dimension $d$, and aim to produce $N$ copies of the state. It is known that this state is drawn from a possible set of states $\Sigma$ with distribution $f(\psi)$ that is normalized as
$$
\int_{\Sigma}f(\psi)d\psi=1.
$$
Dividing the output space $O$ into $N$ qudits labeled $1$ to $N$, our aim is to optimize the quality of the $N$ different copies to be produced. There are different figures of merit that can be applied to cloning (i.e.\ different definitions of the $\mathcal{M}_i$). The simplest figure of merit is the {\em global fidelity}, for which $\mathcal{M}_i=\proj{\psi_i}^{\otimes N}$. The solution to the cloning problem in terms of the global fidelity is known \cite{fiurasek-2001-64} and will not concern us in the rest of the paper. Instead, we consider the {\em single copy fidelity}, in which we take the global output state $\rho_{1\ldots N}$, and assess the fidelity of a single copy on a given site $n$, $F_n=\Tr(\proj{\psi_i}_n\rho_{1\ldots N})$. To this end, we define
$$
\mathcal{M}_i=\sum_{n=1}^N\alpha_n\proj{\psi_i}_n
$$
where imposing $\alpha_n\geq 0$ and $\sum_{n=1}^N\alpha_n=1$ ensures that the $\mathcal{M}_i$ satisfy the required property $\|\mathcal{M}_i\|\leq 1$ in addition to $\mathcal{M}_i \geq 0$. In particular, $F=1$ can still only be achieved if the output state is $\ket{\psi_i}^{\otimes N}$ for all inputs $i$. For generality, we assign different weights $\alpha_n$ to the different copies, emphasizing a possible desire for different qualities of output, although a common desire is equal qualities, $\alpha_n=1/N$. The latter case is called optimal symmetric cloning while the general scenario is the optimal asymmetric cloning problem from $1$ to $N$ copies. 

For the single copy fidelity, on which we henceforth concentrate exclusively, the matrix $R$ is seen to be
\begin{equation}
R=\int_{\Sigma}f(\psi)d\psi\proj{\psi}^T_I\otimes\sum_{n=1}^N\alpha_n\proj{\psi}_n. \label{eqn:RMatrix}
\end{equation}
We are considering here cloning from a single copy of $\ket{\psi}$ so that the problem is $1\rightarrow N$ cloning. This means that the dimension of the input Hilbert space $I$ is no larger than $d$, and the basis is simply $\ket{i}$ for $i=0\ldots d-1$.

\section{State Dependent Cloning of Qubits} \label{sec:qubits}

In this section, we illustrate the above formalism by restricting to the common case of qubits ($d=2$), parameterizing the input state by
$$
\ket{\psi}=\cos\frac{\theta}{2}\ket{0}+\sin\frac{\theta}{2}e^{i\phi}\ket{1}
$$
for some distribution function $f(\theta,\phi)$. We impose two restrictions on the input distribution function $f(\theta,\phi)$, namely: (i) $f(\theta,\phi)$ is phase covariant, meaning it is independent of $\phi$, i.e., $f(\theta,\phi) = f(\theta)$ and (ii) $f(\theta,\phi)$ is symmetric about the equator of the Bloch sphere, i.e., $f(\theta) = f(\pi - \theta)$. 

Using these assumptions\footnote{The assumptions specified have a natural interpretation, but in practice one only needs the following four integrals to vanish: $\int f(\theta,\phi)d\theta d\phi\cos\theta$, $\int f(\theta,\phi)d\theta d\phi e^{i\phi}\sin\theta$, $\int f(\theta,\phi)d\theta d\phi e^{i\phi}\sin(2\theta)$ and $\int f(\theta,\phi)d\theta d\phi e^{i2\phi}\sin^2\theta$.}, the matrix $R$ for the $1 \rightarrow N$ cloning of qubits can be written as
$$
R=\half\identity+\sum_{n=1}^N\alpha_n\Gamma\left(XX-YY+\frac{1-4\Gamma}{2\Gamma}ZZ\right)_{0,n},
$$
where $X$, $Y$ and $Z$ are the familiar Pauli matrices, and
$$
\Gamma=\frac{1}{4}\int f(\theta,\phi)\sin^2\theta d\theta d\phi.
$$
The parameter, $\Gamma$, varying between 0 and $\smallfrac{4}$ provides an intuitive interpretation regarding the distribution of states over the Bloch sphere - the larger the value, the more tightly packed the states are around the equator. We would now like to find the maximum eigenvector and maximum eigenvalue of $R$. The problem can be recast by applying a rotation $Y_0$ to $R$, and instead demanding the minimum eigenvector (ground state) of a new matrix $\tilde{R}$ given as
$$
\tilde R=-\half\identity+\sum_{n=1}^N\alpha_n\Gamma\left(XX+YY+\frac{1-4\Gamma}{2\Gamma}ZZ\right)_{0,n}.
$$
This matrix $\tilde{R}$ is familiar as the Hamiltonian for an anisotropic Heisenberg model on a star configuration and we can proceed to find its ground state.

Before we do that, let us analyze the possibility of implementing the above optimal qubit cloning transformation economically. To do so, we observe that $\tilde{R}$ commutes with the operator $\sum_{n=0}^NZ_n$, so that the system divides into a series of excitation subspaces. We also note that $[\tilde R,X^{\otimes (N+1)}]=0$, which maps the $k$ excitation subspace to the $N+1-k$ excitation subspace. These two observations enable us to make some general statements about the entanglement of the ground state, and hence about the economy of cloning. To do this, consider the minimum eigenvector in an $M$ excitation subspace, which can be split between the input and output systems,
$$
\ket{\Psi_M}=\alpha\ket{0}\ket{\phi_M}+\beta\ket{1}\ket{\phi_{M-1}}.
$$
Since the corresponding state in an $N+1-M$ excitation subspace,
$$
\ket{\Psi_{N+1-M}}=\alpha\ket{1}X^{\otimes N}\ket{\phi_M}+\beta\ket{0}X^{\otimes N}\ket{\phi_{M-1}}
$$
is also an eigenvector, one can construct the state
$$
\frac{1}{\sqrt{2}}(\ket{\Psi_M}+\ket{\Psi_{N+1-M}})
$$
which is also an eigenvector with the same eigenvalue. It is readily verified to be maximally entangled unless $M=\half N, \half(N+1)$ or $\half N+1$. For odd $N$, consider the subspace with $\half(N+1)$ excitations. If $\ket{\Psi_{(N+1)/2}}$ is the unique minimum eigenvector in this subspace, then invariance under $X^{\otimes(N+1)}$ proves that it must be maximally entangled. On the other hand, if this state is not the unique minimum eigenvector, there always exists a state in the degenerate space which is maximally entangled. Hence, for odd $N$, the optimal cloner is always economical, meaning that no ancillary system is required to implement the cloning transformation. For even $N$, this is not necessarily true, but no more than one additional qubit is required because one can increase $N$ by 1 and then implement the optimal cloner, imposing that one of the fidelities is $\half$ ($\alpha_{N+1}=0$). 

\subsection{Symmetric Cloning}

The most commonly studied instance of cloning is where all the output copies are required to have the same fidelity, so $\alpha_n=\frac{1}{N}$. Thus, we have
\begin{eqnarray}
\tilde R=-\half\identity&+&\Gamma\left(X_I\left(\frac{1}{N}\sum_{n=1}^NX_n\right)+Y_I\left(\frac{1}{N}\sum_{n=1}^NY_n\right)\right) \nonumber \\ 
&+&\frac{1-4\Gamma}{2}\left(Z_I\left(\frac{1}{N}\sum_{n=1}^NZ\right)\right), \nonumber
\end{eqnarray}
The permutation invariance on qubits $1\ldots N$ means that the operators such as $\left(\frac{1}{N}\sum_{n=1}^NX_n\right)$ reduce into a simple direct sum structure \footnote{When asymmetry is present, but subsets of spins should have the same fidelity, those sets have that direct sum structure, and are replaced by a single spin operator of the correct dimension.}. In the present instance therefore, the eigenvalue that we require will always be taken from the fully symmetric subspace, meaning that the problem reduces to solving for the minimum eigenvalue of
$$
R'=-\half\identity+\frac{\Gamma}{N}\left(X_IS^{X,N+1}+Y_IS^{Y,N+1}+\Delta Z_IS^{Z,N+1}\right)
$$
where $S^{X,k}$ is the $X$-rotation matrix for a $k$ dimensional space (see the Appendix for further information) and $\Delta=\frac{1-4\Gamma}{2\Gamma}$.
The maximum fidelity that can be realized in the cloning transformation is thus given by
\begin{eqnarray*}
F&=&\half+\frac{\Gamma\Delta}{N} + \\
&& \frac{\Gamma}{N}\max_{i=0\ldots N-1} \sqrt{4(i+1)(N-i)+\Delta^2(N-2i-1)^2}.  
\end{eqnarray*}
The choice of the best $i$ depends on the value of $\Gamma$. For $\smallfrac{6}\leq\Gamma\leq \smallfrac{4}$, the $i$ should ideally be $(N-1)/2$. However, since $i$ must be an integer, it takes the value $\lfloor N/2\rfloor$. Hence, in this regime of $\Gamma$, the fidelity is given by
$$
F=
\left\{\begin{array}{cc}
\half+\frac{1}{2N}+\frac{\Gamma}{N}(N-1) & N \text{ odd} \\
\half+\frac{1-4\Gamma}{2N}+\frac{\Gamma}{N}\sqrt{N(N+2)+\Delta^2} & N \text{ even}
\end{array}\right.
$$
For $0 \leq \Gamma \leq \smallfrac{6}$, the term is maximized when $i$ is either 0 or $N-1$. Both give the same fidelity,
$$
F=\half+\frac{1-4\Gamma}{2N}+\frac{\Gamma}{N}\sqrt{4N+(N-1)^2\Delta^2}.
$$

For classical inputs, when the subset of possible states is only $\ket{0}$ or $\ket{1}$, then it is clear that one should be able to achieve the maximum cloning fidelity 1. This is indeed the case, because when $\Gamma=0$, the maximum eigenstates of $R$ are $\ket{0}^{\otimes (N+1)}$ and $\ket{1}^{\otimes (N+1)}$and one can construct a maximally entangled superposition 
$$
\frac{1}{\sqrt{2}}\left(\ket{0}^{\otimes (N+1)}+\ket{1}^{\otimes (N+1)}\right),
$$
proving that there exists a unitary that achieves the optimal fidelity of cloning. 

If the state $\ket{\psi}$ is selected uniformly over the surface of the Bloch sphere ($\Gamma=\smallfrac{6}$), $R$ becomes the Hamiltonian of the isotropic Heisenberg model. The optimal fidelity of symmetric universal cloning, $F = \frac{2N+1}{3N}$, is recovered.

\subsection{Equatorial Cloning}

In the case of equatorial cloning, the set of states is restricted to the input distribution function $f(\theta)$ with the specific choice of $\theta=\pi/2$. In this situation, we clone states that are on the equator of the Bloch sphere, $\Gamma=\smallfrac{4}$. While we do not solve this problem for general asymmetry, the fully symmetric case \cite{dariano-2003} is now easily recovered by our previous considerations.
$$
F=\left\{\begin{array}{cc}
\frac{3N+1}{4} & N \text{ odd} \\
\half+\frac{\sqrt{N(N+2)}}{4N} & N \text{ even}
\end{array}\right.
$$
We can examine the possibility of achieving the optimal cloning fidelity for equatorial qubit cloning in an economical manner even in the general case of arbitrary asymmetries. Even without completely obtaining the solution, we can make the following statement about the entanglement of the telecloning state -- any optimal pure state must be maximally entangled. This strong statement, showing that the transformation can be performed economically, follows from the observation that $Z_0(\tilde R+\half\identity) Z_0=-(\tilde R+\half\identity)$. Hence, if $\ket{\Psi_M}$ is a minimum eigenvector in a particular excitation subspace, $M$, then $Z_0\ket{\Psi_M}$ is a maximum eigenvector in the same excitation subspace. With the exception of eigenvectors of $\tilde R$ with eigenvalue $-\half$ (which give a fidelity smaller than a complete guess), these two states are orthogonal, making the computational basis the Schmidt basis, with equal Schmidt coefficients. Therefore, we can conclude that for arbitrary asymmetries and any $N$, the optimal equatorial qubit cloner is economical. This generalizes the results in \cite{economy} which show explicitly that for the symmetric case and $N=2$, the equatorial cloner is economical.


Serving as a prelude to the derivation of monogamy relations from solutions to optimal cloning problems in Sec.\ \ref{sec:trades}, we make a brief aside to indicate how the monogamy relation of the CHSH Bell inequality \cite{Toner} arises as a consequence of the consideration of $1 \rightarrow 2$ asymmetric equatorial cloning. The CHSH inequality is a two-party two-setting Bell inequality. Up to local unitaries, the operator measured in a CHSH test on a quantum system between parties 0 and $n$ can be expressed as
$$
\mathcal{B}_{0n}=\sqrt{2}(XX-YY)_{0n}.
$$
This is due to the fact that (up to local unitaries), the individual settings can be chosen to lie in the X-Y plane by both parties. Moreover, for optimal violation for any entangled quantum state, the local settings should be chosen such that they are as far away from being commuting as possible \cite{Tsirelson}. In this scenario, we would like to derive the optimal trade-off between the CHSH inequality violation between Alice and Bob (each holding a qubit state) and that between Alice and Charlie, i.e., between $\mathcal{B}_{01}$ and $\mathcal{B}_{02}$. Maximizing the cloning fidelity corresponds to finding the maximum eigenvalue, $\lambda$, of
$$
\half\identity+\frac{1}{4\sqrt{2}}\left(\alpha_1\mathcal{B}_{01}+\alpha_2\mathcal{B}_{02}\right),
$$
which implies that the average values of the two CHSH parameters for any state, $\langle\mathcal{B}_{01}\rangle$ and $\langle\mathcal{B}_{02}\rangle$, obey
$$
\alpha_1\langle\mathcal{B}_{01}\rangle+\alpha_2\langle\mathcal{B}_{02}\rangle\leq 4\sqrt{2}(\lambda-\half)
$$
where the maximum eigenvalue is given by $\lambda=\half(\sqrt{\alpha_1^2+\alpha_2^2}+1)$. The two asymmetry parameters must satisfy $\alpha_1+\alpha_2=1$, so we can choose them to satisfy 
$$
\alpha_1=\langle\mathcal{B}_{01}\rangle\kappa \qquad \alpha_2=\langle\mathcal{B}_{02}\rangle\kappa,
$$
for suitable $\kappa$. This choice of the asymmetry parameters implies the well-known monogamy relation for the Bell-CHSH inequality within quantum theory \cite{Toner}
$$
\langle\mathcal{B}_{01}\rangle^2+\langle\mathcal{B}_{02}\rangle^2\leq 8.
$$
In general, when interpreting the state $\ket{\Psi}$ as a state used in telecloning, its ability to clone well (and hence the general performance of the corresponding unitary cloner) is limited by the monogamy of its correlations. This could be derived for any of the state dependent cloners. Doing so for the universal cloner produced singlet monogamy \cite{our_paper} (see Sec.\ \ref{sec:trades}).


\section{$1 \rightarrow N$ Qudit Cloning} \label{sec:main}

We shall now move to the main result of the paper, namely the solution to the general case of $1 \rightarrow N$ cloning where the input state lives in an arbitrary Hilbert space dimension $d$, and the output copies can have differing fidelities. For this qudit cloning problem, we concentrate on the specific case when $f(\psi)$ is uniformly distributed over the set of states according to the Haar measure i.e.\ we are equally likely to be trying to clone any one qudit pure state. To derive the matrix $R$ for this, the $1 \rightarrow N$ universal asymmetric qudit cloning, problem, we write that $\ket{\psi}=U\ket{0}$ and integrate over $U$ to obtain \cite{our_paper}
$$
R=d\int dUU^*\proj{0}U^T\otimes \sum_{n=1}^N\alpha_nU^\dagger\proj{0}_nU.
$$
The integration results in twirling \cite{horo_twirl} to yield
$$
R=\frac{1}{d+1}\identity+\frac{d}{d+1}\sum_{n=1}^N\alpha_n\proj{B}_{0,n}.
$$
where $\ket{B}=\frac{1}{\sqrt{d}}\sum_{n=0}^{d-1}\ket{\phi_i}_I\ket{\phi_i}_O$ is the Bell state as defined previously.

The main route towards solving this is to notice that any symmetric state $\ket{\Phi}$ of $N-1$ qudits (meaning the same state is returned when any pair of qudits is exchanged) allows one to write down states $\{\ket{\Psi_m}=\ket{B}_{0,m}\ket{\Phi^{N-1}}_{1,2\ldots N\neq m}\}$ that span a subspace of $R$. The permutation invariance ensures that
$$
\proj{B}_{0,m}\ket{\Psi_n}=\frac{1+(d-1)\delta_{n,m}}{d}\ket{\Psi_m}.
$$
Within the subspace, the eigenvectors of the matrix $R$ are described by
\begin{equation}
\ket{\Psi}=\sum_{n=1}^N\beta_n\ket{\Psi_n},	\label{eqn:state_form}
\end{equation}
subject to the normalization condition
\begin{equation}
\left(\sum_{n=1}^N\beta_n\right)^2+(d-1)\sum_{n=1}^N\beta_n^2=d. \label{eqn:norm}
\end{equation}
Here the coefficients $\{\beta_n\}$ are chosen such that they form a vector which is the maximum eigenvector of the matrix
\begin{equation}
\sum_{n,m=1}^N\alpha_n\ket{n}\bra{m}+(d-1)\sum_{n=1}^N\alpha_n\proj{n}.
\label{eqn:big_result}
\end{equation}
By definition, then, this is the maximum eigenvector of the particular subspace, eigenvalue $\lambda$. If it were the maximum eigenvector of the entire matrix $R$, the cloning fidelity would be given by $F\leq\frac{1+\lambda}{d+1}$. To prove that this eigenvector is indeed the maximum eigenvector, we invoke a variant of the Lieb-Mattis Theorem \cite{liebmattis}. Indeed, if $d=2$, no modification is required, and \cite{liebmattis} instantly proves that we have the maximum eigenvector because, by the previous results, we are searching for the ground state of a Heisenberg star. The full proof for arbitrary $d$ is deferred to the Appendix.

It should be noted that the matrix in Eq.\ (\ref{eqn:big_result}), which is derived from the matrix $R$ and used to choose the particular set of $\{\beta_n\}$, is independent of the choice of $\ket{\Phi}$. Also note that since the matrix in Eq.\ (\ref{eqn:big_result}) is real, the coefficients $\beta_n$ can be taken to be real, justifying the absence of conjugates in the normalisation condition. 

\subsection{Possibility of economic universal qudit cloning}

Having proved, as in the Appendix, that the state Eq.\ (\ref{eqn:state_form}) gives the maximum eigenvector of $R$ and that the optimal fidelity is given by $F\leq\frac{1+\lambda}{d+1}$, we need to check that this bound can be saturated by verifying that the maximum eigenvector corresponds to an implementable CP map. To show this, we need to demonstrate that there exists a superposition or mixture of the set of optimal states such that the partial trace over the outputs returns a maximally mixed state on the input (Sec.\ \ref{sec:isomorphism}).

Eq.\ (\ref{eqn:state_form}) offers plenty of freedom in selecting the symmetric state $\ket{\Phi}$. Hence, we will pick a particular symmetric state
$$
\ket{\Phi}=\frac{1}{\sqrt{d}}\sum_{i=0}^{d-1}\ket{i}^{\otimes (N-1)}.
$$
Provided $N>2$, tracing out qudits 1 to $N$ of the corresponding $\ket{\Psi}$ just leaves
$$
\rho_0=\smallfrac{d}\identity,
$$
i.e.\ $\ket{\Psi}$ is maximally entangled, so the optimal map can be implemented, and that implementation is economical.

On the other hand, for $N=2$, one can examine an arbitrary $\ket{\Phi}$ and prove that the maximally entangled state only occurs if $\beta_1\beta_2=0$ i.e.\ one of the clones will be perfect, and the other has no information about the input qudit at all; this is the trivial $N=1$ case. Otherwise, economical optimal cloning if impossible for $N=2$. Nevertheless, one can readily verify that the mixed state formed by the maximum eigenvectors
$$
\rho=\frac{1}{d}\sum_{i=0}^{d-1}\proj{\Psi_i}
$$
has the required partial trace $\rho_0 = \smallfrac{d} \identity$, meaning that the optimal fidelity can be realized with only one ancilla of dimension $d$. The results of \cite{economy}, where it was proven that universal economical cloning is impossible for the $1 \rightarrow 2$ case, are clearly a special case of this result.

\section{$M \rightarrow N$ Qubit Cloning} \label{sec:many}

The study of the optimal cloning machines can be extended to the case when, instead of a single input copy of the state to be cloned, multiple copies are provided. The matrix $R$ can be derived for this scenario in a similar manner to the previous sections. The difference here is that the dimension $d'$ of the input space is not simply $d^M$. Instead, $d'$ is the dimension of the fully symmetric subspace of the $M$ qudits. The matrix $R$ is therefore defined as
$$
R=d'\int f(\psi)d\psi\bigotimes_{m=1}^N\proj{\psi}^T_{I_m}\sum_{n=1}^N\alpha_n\proj{\psi}_n.
$$
As before, the maximum eigenvalue of $R$ upper bounds the achievable fidelity of the cloning transformation and, if there exists some superposition of the maximum eigenvectors which is maximally bipartite entangled between the input qudits and the output ones, the maximum fidelity can be achieved by a unitary operation. 

Here, we concentrate on the special case of universal $M \rightarrow N$ qubit cloning. After performing the twirling operation and projecting the input spins onto the symmetric subspace, we obtain
$$
R=\sum_n\alpha_n R_n
$$
with
$$
2R_n=\identity+\frac{1}{M+2}(S^{X,M+1}_IX_n-S^{Y,M+1}_IY_n+S^{Z,M+1}_IZ_n).
$$
The fidelities of the output copies are then given by $F_n=\bra{\Psi}R_n\ket{\Psi}$ with the final fidelity being $F=\sum_n\alpha_nF_n$.

The proof strategy exactly mirrors that of the $1 \rightarrow N$ universal cloning task (see the Appendix). We first observe that the set of states
$$
\ket{\psi_{x,i}}=\frac{1}{\sqrt{M+1}}\sum_{n=0}^{M}\ket{n}_I\ket{\Phi_n^M}_{\{x_k=1\}}\ket{\Phi_{i}^{N-M}}_{\{x_k=0\}}
$$
forms a closed subspace. Here $\ket{n}_I$ represents the basis elements for the input symmetric subspace and $x\in\{0,1\}^N$ is an $N$-bit string of weight $M$. The superposition above therefore denotes that a symmetric state $\ket{\Phi_n^M}$ is the state of the $M$ output spins for which the bit value in the bit string is $1$ (of which $n$ are in the state $\ket{1}$ and $M-n$ are $\ket{0}$). $\ket{\Phi_{i}^{N-M}}$ is the symmetric state of the remaining $N-M$ output spins with $i$ qubits in the state $\ket{1}$. To verify that these states form a closed subspace of $R$, note that
$$
R_n\ket{\psi_{x,i}}=\left\{\begin{array}{cc}
\ket{\psi_{x,i}} & x_n=1\\
\frac{1}{M+2}\left(\ket{\psi_{x,i}}+\sum_y\ket{\psi_{y,i}}\right) & x_n=0
\end{array}\right.
$$
Here the sum over $y$ is over all bit strings $y\in\{0,1\}^N$ of weight $M$ with the condition that $y=x$ in all except two places, one of which is bit $n$, $y_n=1$. The relation
$$
\braket{\psi_{x,i}}{\psi_{z,i}}=\frac{1}{M+1-x\cdot z}			
$$
then implies the normalization condition.

Since the $\{\ket{\psi_{x,i}}\}$ span the space, we can write  
\begin{equation}
\ket{\Psi_i}=\sum_x\beta_x\ket{\psi_{x,i}}. \label{eqn:m_n_qubit}
\end{equation}
The matrix $R$ within this restricted subspace, $\tilde R$, is written, for bit strings $x,y\in\{0,1\}^N$ of weight $M$, as
$$
\tilde R_{x,y}= \left\{\begin{array}{cc}
\frac{1}{M+2}+\frac{M+1}{M+2}\sum_{n:x_n=1}\alpha_n & x=y \\
\frac{1}{M+2}\sum_{n:x_n=0}\alpha_n\delta_{y_n=1} & x\cdot y=M-1
\end{array}\right. 
$$
So, at least in principle, the maximum eigenvalue of this matrix can be found.  The Lieb-Mattis theorem is directly applicable, so we prove that the subspace contains the maximum eigenvector for the entire $R$, revealing the upper bound on the optimal cloning fidelity. It still remains to be checked that the optimal fidelity can be implemented by a CP map, which is more readily achieved on a case-by-case basis.

\section{Fidelity Trade-off Relations} \label{sec:trades}

The formalism outlined so far reduces the task of finding the optimal universal qubit cloner to that of diagonalizing an $\binom{N}{M}\times\binom{N}{M}$ matrix, although we have not explicitly performed this diagonalization in the asymmetric case. Finding the maximal eigenvalue specifies a trade-off between the parameters $\{\alpha_n\}$ and the qualities of the clones given by $F_n=\bra{\Psi}R_n\ket{\Psi}$. Evidently, specifying the solution to the asymmetric cloning problem is best done in terms of trade-off relations between the various output fidelities. In this section, we derive the fidelity trade-off relations in two cases of interest to us, namely $1 \rightarrow N$ universal qudit cloning and $(N-1) \rightarrow N$ universal qubit cloning. 

Whenever the task of finding the maximum eigenvector of $R$ can be reduced to a problem on an $N$-dimensional space (where $N$ is the number of different fidelities we might have in the asymmetric cloning), we have $N$ parameters $\beta_i$ in the state
$$
\ket{\Psi}=\sum_{n=1}^N\beta_i\ket{\psi_i}
$$
where $\{\ket{\psi_i}\}$ span the space. The fidelities $F_n$ are simply a quadratic function of the $\beta_i$, as is the normalization condition $\braket{\Psi}{\Psi}=1$. We can therefore expect to eliminate the $\beta_i$ in the normalization condition and replace them with the $F_n$. Intuitively, one can understand that when the dimension of the space is equal to the number of asymmetry parameters $\alpha_n$, it is possible to explore the entire subspace corresponding to the optimal solution by varying the different parameters $\alpha_n$. If the dimension of the subspace is greater, this is certainly not possible, and we can only follow a constrained path within the space, as specified by the maximum eigenvector. At that point, extracting the optimal trade-off relation analytically becomes challenging. Fortunately, in the two particular cases specified previously, it is possible to derive these fidelity relationships by the method outlined above. It is important to note that neither case ever actually requires the determination of the maximum eigenvalue of $R$.

\subsection{Trade-offs for the $1 \rightarrow N$ qudit cloner}

For $1\rightarrow N$ universal qudit cloning, we derived the normalization condition of Eq.\ (\ref{eqn:norm}). This means that a (possibly non-optimal state) must satisfy the inequality
$$
\left(\sum_{n=1}^N\beta_n\right)^2+(d-1)\sum_{n=1}^N\beta_n^2\leq d.
$$
The cloning fidelities are given by
$$
F_n=\frac{1}{d+1}+\frac{1}{d(d+1)}\left((d-1)\beta_n+\sum_{m=1}^N\beta_m\right)^2.
$$
In order to eliminate the $\beta_n$ in favor of the cloning fidelities $F_n$, observe that
$$
\sum_{n=1}^N\sqrt{d(d+1)F_n-d}=(d-1+N)\sum_{n=1}^N\beta_n.
$$
For simplicity, the trade-off relations can be expressed in terms of the singlet fraction, $p_{0,n}$. The singlet fraction of a pair of states between $0$ and $n$ is defined as
$$
p_{0,n}=\max_{U,V}\bra{B}U\otimes V\rho_{0,n}U^\dagger\otimes V^\dagger\ket{B}
$$
where $U$ and $V$ are arbitrary $d$-dimensional unitary rotations. It is directly related to the cloning fidelity in $1\rightarrow N$ universal qudit cloning \cite{horo_teleport},
$$
F_n=\frac{p_{0,n}d+1}{d+1}.
$$
The previous expression can now be rewritten in terms of the singlet fraction as
$$
d\sum_{n=1}^N\sqrt{p_{0,n}}=(d-1+N)\sum_{n=1}^N\beta_n.
$$
Eliminating the variables $\beta_n$ from the normalization condition, the trade-off relation in terms of the singlet fractions for the $1 \rightarrow N$ universal asymmetric qudit cloner is obtained:
\begin{equation}
\sum_{n=1}^Np_{0,n}\leq\frac{d-1}{d}+\frac{1}{N+d-1}\left(\sum_{n=1}^N\sqrt{p_{0,n}}\right)^2.
\label{eqn:new_monogamy}
\end{equation}
This relation describes the `singlet monogamy' \cite{our_paper}, in comparison to the tangle monogamy \cite{tangle1} or Bell monogamy \cite{Toner}, as it serves to bound the correlations that are achievable between different sets of spins in any quantum state. It can be seen that this single equation encapsulates the known results on $1\rightarrow N$ universal asymmetric cloning, subsuming previous results on $1\rightarrow 4$ cloning \cite{asymmetric,subsume}, and much more besides.

\subsection{Trade-offs for the $(N-1) \rightarrow N$ qubit cloner}

Aside from $M=1$ universal cloning, another instance for which the fidelity trade-off relation can be derived is $(N-1)\rightarrow N$ universal asymmetric qubit cloning. Let us replace the $\beta_x$ that appear in the solution to this problem in Eq.\ (\ref{eqn:m_n_qubit}) with $\gamma_n$, where, since the bit string $x$ is of length $N$ and weight $N-1$, there is only one zero bit, which we denote as being at position $n$. For this problem, the normalization condition reads
$$
2\geq\left(\sum_{n=1}^{N}\gamma_n\right)^2+\sum_{n=1}^{N}\gamma_n^2,
$$
while each of the cloning fidelities can be found to be
$$
F_n=1-\frac{\gamma_n^2}{2}.
$$
The elimination of the $\gamma_n$ therefore leads to the following fidelity trade-off relation for the case of $(N-1) \rightarrow N$ universal asymmetric qubit cloning 
\begin{equation}
N-1\leq\sum_{n=1}^{N}F_n-\left(\sum_{n=1}^{N}\sqrt{1-F_n}\right)^2,
\label{eqn:Nmonogamy}
\end{equation}
describing the entire set of achievable fidelities. 

Finally, to verify that these fidelities can be achieved by a CP map, observe that an equal mixture of the two states $\ket{\Psi_0}$ and $\ket{\Psi_{1}}$, as in Eq.\ (\ref{eqn:m_n_qubit}), yields the identity on the $N-1$ input spins when the $N$ output spins are traced out, 
$$
\half\Tr_{1\ldots N}(\proj{\Psi_{0}}+\proj{\Psi_{1}})=\smallfrac{N}\identity.
$$
This implies that this cloning transformation uses one ancilla qubit (i.e.\ is not economical) but gives the requisite trace satisfying the necessary and sufficient condition for its implementation by a CP map. 

\section{Conclusions}
The efficacy with which a given quantum state can be cloned is an important problem in quantum information, having implications in diverse fields such as quantum cryptography, state estimation, and entanglement monogamy. In this paper, we tackled the problem using the Choi-Jamio\l kowski isomorphism, and provided solutions in a number of interesting scenarios. The investigation of $1 \rightarrow N$ qubit cloning yielded a parametrization for a large class of state dependent cloners including the equatorial and universal cloners. If $N$ is odd, the optimal cloner is economical and when $N$ is even, no more than one ancilla qubit is required. As an interesting aside, we derived the monogamy relation for the CHSH Bell inequalities in terms of the fidelity trade-offs in the $1$ to $2$ asymmetric equatorial qubit cloning problem. For the general $1$ to $N$ asymmetric equatorial cloning problem, we proved that for any $N$ and any asymmetry, the cloner is economical. 

The main result of the paper is the solution to the $1 \rightarrow N$ universal asymmetric qudit cloning problem, for which we derived the exact trade-off relations in the fidelities of the output clones, Eq.\ (\ref{eqn:new_monogamy}). For $N>2$, the optimal cloner is economical while for $N = 2$, it is not, requiring one additional ancillary qudit for its implementation. Generalizing to $M \rightarrow N$ cloning, we derived the (tight) trade-offs in the fidelities for the special case of $(N-1)$ to $N$ universal cloning of qubits, Eq.\ (\ref{eqn:Nmonogamy}), and provided some insight into why it is these two cases that yield such a straightforward monogamy relationship.

One aspect that we have not touched upon here is the efficiency of implementation of the optimal cloners. However, states such as those of the $1\rightarrow N$ cloner, Eq.\ (\ref{eqn:state_form}), have a concise description in the computational basis, which is sufficient to guarantee that they can be easily prepared, i.e.\ with circuits that scale as a polynomial in $N$ \cite{kaye}. While such an argument says nothing about the economy of the implementation, as ancillas may be used in intermediate steps, economy can also be achieved.

The formalism presented here, and, primarily, the techniques for proving optimality, can potentially be applied in many other scenarios. A natural generalization involves the cloning of mixed quantum states, a problem known as broadcasting \cite{broadcasting}. There has also been a vast amount of literature on the problem of cloning for continuous variable systems \cite{cv-systems} to which the methods may be directly applied. Finally, it is potentially interesting to consider the optimal cloning of quantum observables and entanglement rather than entire quantum states \cite{clone-ent}.

{\em Acknowledgements:} This work was supported by the National Research Foundation \& Ministry of Education, Singapore.

\newpage
\appendix
\section{Derivation of Optimal Universal Qudit Cloning}

In this Appendix, we provide the optimal solution to the $1 \rightarrow N$ universal qudit cloning problem. In particular, we find the maximum eigenvector of the matrix $R$ which is defined for this problem as
$$
R=\frac{1}{d+1}\identity+\frac{d}{d+1}\sum_{n=1}^N\alpha_n\proj{B}_{0,n}.
$$

A major ingredient in the derivation of our results is the understanding of symmetries in a system, particularly those of rotational and permutation invariance. We will recap the bare minimum here in order to familiarize the reader with our notation, but for further details, the excellent treatment in \cite{frames} can be relied upon.
We start by defining the spin operators
\begin{eqnarray*}
S^{X,d}&=&\sum_{n=0}^{d-2}\sqrt{(n+1)(d-n-1)}(\ket{n}\bra{n+1}+\ket{n+1}\bra{n})	\\
S^{Y,d}&=&\sum_{n=0}^{d-2}\sqrt{(n+1)(d-n-1)}(-i\ket{n}\bra{n+1}+i\ket{n+1}\bra{n})	\\
S^{Z,d}&=&\sum_{n=0}^{d-1}(d-1-2n)\proj{n},
\end{eqnarray*}
where the as yet unused subscript will be used later to denote which spin the operators act on. If $d=2$, these operators will be simply written as $X$, $Y$ and $Z$ respectively, recovering the usual Pauli operators. In the $1 \rightarrow N$ cloning problem, we have one input spin ($I$) of dimension $d'$, and $N$ output spins ($1\ldots N$) of dimension $d$. These will require us to specify the total spin operators
\begin{eqnarray*}
J_X^{d',d}&=&-S_I^{X,d'}+\sum_{n=1}^NS_n^{X,d} \\
J_Y^{d',d}&=&S_I^{Y,d'}+\sum_{n=1}^NS_n^{Y,d} \\
J_Z^{d',d}&=&-S_I^{Z,d'}+\sum_{n=1}^NS_n^{Z,d} \\
J^2&=&J_X^2+J_Y^2+J_Z^2
\end{eqnarray*}
Note that we have an unusual combination of negative signs in these definitions. This is due to the symmetry structure not being the usual invariance under the unitaries $U^{\otimes N+1}$. However, note that under the action of
$$
U_I=\sum_{n=0}^{d-1}(-1)^n\ket{n}\bra{d-1-n}
$$
on the input spin, this is transformed into the usual total spin operators. For this reason, we instantly recover the commutation relation $[J_Z,J^2]=0$. Since $R$ simultaneously commutes with both $J^2$ and $J_Z$, it has two quantum numbers $S$ (where $4S(S+1)$ is an eigenvalue of $J^2$) and $M_Z$ (taking values $-S$ to $S$ in integer steps), providing a lot of information about its structure.  

We now use the permutation invariant states of $M$ qudits of fixed excitation number $k = 0 \ldots M(d-1)$ which are denoted as $\ket{\Phi^M_k}$. In the case of $d=2$, these are the uniform superposition of all states of $M$ qubits containing $k$ excitations. For a set of $M$ $d$-dimensional spins, we note their relations to the spin operators,
\begin{widetext}
\begin{eqnarray*}
\left(\sum_{m=1}^MS_m^{Z,d}\right)\ket{\Phi^M_{k}}&=&(M(d-1)-2k)\ket{\Phi^M_{k}} \\
\half\left(\sum_{m=1}^MS_m^{X,d}+iS_m^{Y,d}\right)\ket{\Phi^M_{k}}&=&\sqrt{k(M(d-1)+1-k)}\ket{\Phi^M_{k-1}} \\
\half\left(\sum_{m=1}^MS_m^{X,d}-iS_m^{Y,d}\right)\ket{\Phi^M_{k}}&=&\sqrt{(k+1)(M(d-1)-k)}\ket{\Phi^M_{k+1}}.
\end{eqnarray*}
Under the action of the individual terms in the matrix $R$, these states transform as
$$
\proj{B}_{0,n}\ket{B}_{0,m}\ket{\Phi_i^{N-1}}_{1,2\ldots N\neq m}=\left\{\begin{array}{cc}
\ket{B}_{0,n}\ket{\Phi_i^{N-1}}_{1,2\ldots N\neq n} & n=m \\
\frac{1}{d}\ket{B}_{0,n}\ket{\Phi_i^{N-1}}_{1,2\ldots N\neq n} & n\neq m
\end{array}\right.
$$
\end{widetext}
This therefore implies that the set of states $\{\ket{B}_{0,m}\ket{\Phi_i^{N-1}}_{1,2\ldots N\neq m}\}$ spans a subspace of $R$ for each $i$, corresponding to a value of $M_Z=\half(d-1)(N-1)-i$. Within this subspace, the eigenvectors are given by a superposition 
\begin{equation}
\ket{\Psi_i}=\sum_{n=1}^N\beta_n\ket{B}_{0,n}\ket{\Phi_i^{N-1}}_{1,2\ldots N\neq n},	\label{eqn:state}
\end{equation}
and the coefficients must satisfy the normalization condition
$$
\left(\sum_{n=1}^N\beta_n\right)^2+(d-1)\sum_{n=1}^N\beta_n^2=d.
$$
Moreover, the coefficients $\{\beta_n\}$ are chosen to be the maximum eigenvector of the matrix
\begin{equation}
\sum_{n,m=1}^N\alpha_n\ket{n}\bra{m}+(d-1)\sum_{n=1}^N\alpha_n\proj{n}, \label{eqn:result}
\end{equation}
this being independent of $i$. The state Eq. (\ref{eqn:state}) formed using the coefficients $\beta_n$ from the maximum eigenvector of the above matrix, is the maximum eigenvector of $R$ within a given subspace $i$. The cloning fidelity using the maximum eigenvector is then given in terms of the maximum eigenvalue $\lambda$ as $F=\frac{1+\lambda}{d+1}$. Before we proceed to find the absolute maximum eigenvector of $R$ over all subspaces we examine the known case of symmetric cloning.

\subsection{Symmetric Cloning}

For the symmetric cloner where each of the output clones in the $1 \rightarrow N$ universal qudit cloner have the same quality so that $\alpha_n=\smallfrac{N}$, we can readily proceed to calculate the $\beta_n$ and $F$. All the $\beta_n$ are equal, and hence, by the normalization condition, we have
$$
\beta_n=\sqrt{\frac{d}{N^2+N(d-1)}}
$$
and the final fidelity is
$$
F=\frac{1}{N}+\frac{2(N-1)}{N(d+1)}.
$$
This is known to be the optimal fidelity of the symmetric universal qudit cloner \cite{rmp}, so even without finding any other eigenvalues, we infer that these eigenvectors (given by Eq. (\ref{eqn:state}) with equal $\beta_n$) must be the maximal ones. These states, which we denote by $\ket{\Psi_{i}^{\text{sym}}}$, will be used later in the proof.

Let us make an important observation about these states. The eigenvectors $\ket{\Psi_{i}^{\text{sym}}}$ within a particular $M_Z$ subspace can be chosen to have a positive amplitude on every basis state except for those special cases in which the qudit state on the input spin 0 does not appear in the rest of the string. For these special cases, such as the state $\ket{31210}$, which does not have a $\ket{3}$ on the any of the sites except spin 0, the amplitude of the eigenvector $\ket{\Psi_{i}^{\text{sym}}}$ is $0$. These special cases can be ignored because they are eigenvectors of $R$ with eigenvalue $1/(d+1)$ (for any arbitrary set of $\{\alpha_n\}$), and are hence never optimal. 
It is important to note that not every choice of $\ket{\Phi_i^{N-1}}$ has this property of having a positive amplitude on every basis state within the particular subspace but one can always choose the eigenvectors $\ket{\Psi_{i}^{\text{sym}}}$ to have this property. For example, choices of $\ket{\Phi_2^2}$ could be $\ket{11}$ or $(\ket{02}+\ket{20})/\sqrt{2}$ for $d\geq 3$, and these do not have positive amplitude on every basis state of the relevant $M_Z$, but there are superpositions which do. 

\subsection{Asymmetric Cloning}

It is now our task to prove that the maximum eigenvector $\ket{\Psi_i}$ given in Eq.\ (\ref{eqn:state}) within this subspace of given $i$ is always the maximum eigenvector of $R$, for arbitrary asymmetries. To do so, we are going to make use of a modified Lieb-Mattis theorem \cite{liebmattis}. The idea behind the proof is the following: we will prove that the maximum eigenvector within a given $M_Z$ subspace (making no restriction on $S$) has non-negative coefficients on all the basis states. Since (apart from the discounted special cases) $\ket{\Psi_i^{\text{sym}}}$ has positive amplitudes on all these basis states, this implies that
$$
\braket{\Psi_i}{\Psi_i^{\text{sym}}}>0.
$$
Therefore, the $\ket{\Psi_i}$ must be drawn from the subspace $\{\ket{B}_{0,m}\ket{\Phi_i^{N-1}}_{1,2\ldots N\neq m}\}$, and hence must be the maximum eigenvector that we solve for in Eq.\ (\ref{eqn:result}).

This would successfully establish that for the subspaces $M_Z=-\half(d-1)(N-1),\ldots ,\half(d-1)(N-1)$, we know the maximum eigenvector. The maximum eigenvector in the other subspaces cannot have a larger eigenvalue due to the following argument. If it were true that these other subspaces had an eigenvector with a larger eigenvalue, then this eigenvector must be drawn from a subspace of $S>\half(d-1)(N-1)$. However, any such eigenvalue is degenerate in $M_Z$ so it would also be present in all other $M_Z$ subspaces from $-S$ to $S$, in particular in the ones in which we have already found a different maximum eigenvector. So, by contradiction, this establishes that we have the maximum eigenvalue, $F=(1+\lambda)/(d+1)$, and that it is (at least) $(d-1)(N-1)+1$ fold degenerate.

This leaves us with the task of proving that, within a given $M_Z$ subspace, the coefficients of the maximum eigenvector are non-negative. To do so, divide $R$ into two components, $R_0$ and $R_1$, where $R_0$ contains all the diagonal elements, and $R_1$ is the remainder. In the computational basis, $\ket{a}$, we have $\bra{a}R_0\ket{a}=e_a$ and $\bra{a}R_1\ket{b}=K_{ab}$. Note that $K_{ab}\geq 0$. Assume that in a particular excitation subspace, $M_Z$, we know the maximum eigenvector,
$$
\ket{\psi}=\sum_a f_a\ket{a},
$$
with eigenvalue $E_{M_Z}$. Hence,
$$
\sum_b K_{ba}f_b=(E_{M_Z}-e_a)f_a.
$$
Any other state must have a smaller expectation value of $R$, unless it is also a maximum eigenvector. Let us first try a state $\ket{a}$ as the ansatz. This reveals $e_a\leq E_{M_Z}$. Hence, we can take the modulus of the above equation,
$$
\left|\sum_bK_{ba}f_b\right|=(E_{M_Z}-e_a)|f_a|.
$$
Next, we try the ansatz state
$$
\ket{\tilde\psi}=\sum_a|f_a|\ket{a}.
$$
This requires that there is at least one non-zero $f_a$ such that
$$
\sum_bK_{ba}|f_b|\leq(E_{M_Z}-e_a)|f_a|
$$
but since $\sum_bK_{ba}|f_b|\geq\left|\sum_bK_{ba}f_b\right|$, this can only be satisfied with equality for every $a$, meaning that the coefficients of the maximum eigenvector in each $M_Z$ subspace satisfy
$$
f_a\geq 0,
$$
up to a global phase factor. This completes the proof that the maximum eigenvector within a given $M_Z$ subspace has non-negative coefficients on all the basis states. By the arguments outlined previously, this establishes that the state Eq.\ (\ref{eqn:state}) is the maximum eigenvector of $R$, the cloning fidelity being given in terms of the maximum eigenvalue $\lambda$ by $F=\frac{1+\lambda}{d+1}$. 

\end{document}